\documentstyle[aps,epsfig]{revtex}
\begin{document}
\draft
\title{Interplay   of  friction  and  noise  and  enhancement  of
disoriented chiral condensate}
\author{\bf A. K. Chaudhuri \cite{byline}}
\address{
Variable Energy Cyclotron Centre\\
1-AF, Bidhan Nagar, Calcutta- 700 064}
\maketitle
\begin{abstract}
Using  the  Langevin  equation  for the linear $\sigma$ model, we
have investigated the effect of  friction  and  noise  on  the
possible  disoriented  chiral  condensate  formation. Friction and
noise are supposed to suppress  longwavelength  oscillations  and
growth   of   disoriented   chiral  condensate  domains.  Details
simulation shows that for  heavy  ion  collisions,  interplay  of
friction  and  noise  occur  in  such  a manner that formation of
disoriented chiral condensate domains are enhanced.

\end{abstract}


In recent years there is much excitement about the possibility of
formation  of  disoriented  chiral condensate (DCC). The idea was
suggested by  Rajagopal  and  Wilczek  \cite{ra93}  and  also  by
Bjorken  and others \cite{bj93}. In hadron-hadron or in heavy ion
collisions, a macroscopic region of  space-time  may  be  created
within  which  the  chiral order parameter is not oriented in the
same direction in the internal $O(4)=SU(2) \times SU(2)$ space as
the ordinary vacuum. The misaligned condensate has the same quark
content  and  quantum  numbers  as  do  pions   and   essentially
constitute  a  classical  pion  field.  The  system  will finally
relaxes to the true vacuum and in the process can  emit  coherent
pions.  Possibility  of  producing classical pion fields in heavy
ion collisions had been discussed earlier by Anslem  \cite{an91}.
DCC  formation in hadronic or in heavy ion collisions can lead to
the spectacular events that some portion of the detector will  be
dominated by charged pions or by neutral pions only. In contrast,
in  a  general  event  all  the  three pions ($\pi^+$,$\pi^-$ and
$\pi^0$) will be equally well produced.  This  may  then  be  the
natural explanation of the so called Centauro events \cite{la80}.

Microscopic  physics governing DCC phenomena is not well known. It
is in the regime of non-perturbative QCD  as  well  as  nonlinear
phenomena,   theoretical  understanding  of  both  of  which  are
limited. One thus uses some effective field  theory  like  linear
$\sigma$ model with various approximations to simulate the chiral
phase transition \cite{rand,gavi,asak}. In the linear sigma model
chiral  degrees  of  freedom  are  described by the the real O(4)
field $\Phi=(\sigma,\roarrow{\pi})$. Because of  the  isomorphism
between  the  groups  $O(4)$  and $SU(2) \times SU(2)$, the later
being the appropriate group for  two  flavour  QCD,  linear  sigma
model  can  effectively  model  the  low  energy  dynamic  of QCD
\cite{bo96}. Explicit simulation with linear sigma model, indicate
that DCC depends critically on the initial field  configurations.
With  quench like initial condition DCC domains of 4-5 fm in size
can form \cite{asak}. Initial conditions other than  quench  lead
to  much  smaller  domain  size.  Quench scenario assume that the
effective potential governing the evolution of long  wave  length
modes  immediately  after  the phase transition at $T_c$ turns to
classical one at zero temperature. It can happen only in case  of
very  rapid  cooling  and expansion of the fireball. In heavy ion
collisions quench like initial conditions are unlikely.

Very  recently, effect of external media on possible DCC is being
investigated   \cite{gr97,bi97,zu00,ri98,ch99a,ch99b,ch99c,ch00}.
Indeed,  in  heavy ion collisions, even if some region is created
where   chiral   symmetry   is  restored,  that  region  will  be
continually interacting with surrounding medium  (mostly  pions).
The  surrounding  medium  can be conveniently modelled by a white
noise source  and  one  can  use  Langevin  equation  for  linear
$\sigma$  model to simulate the DCC formation under the influence
of external media. Recently it has been shown  that  in  $\phi^4$
model, hard modes can be integrated out to obtain a Langevin type
of  equations  for  the  soft modes \cite{gr97}. Biro and Greiner
\cite{bi97} using a Langevin equation  for  the  linear  $\sigma$
model,  investigated the interplay of friction and white noise on
the evolution and stability of {\em zero mode}  pion  fields.  In
general  friction  and  noise  reduces  the amplification of zero
modes. But in some trajectories, large  amplification  may  occur
\cite{bi97}. We also obtain similar results \cite{ch99a}.

It is popular expectation that friction and noise will reduce the
DCC  formation  probability. The expectation was found to be true
in  one   dimensional   calculations   with   zero   modes   only
\cite{bi97,ch99a}.  To see how far this expectation is valid when
higher  modes  are  included,  in  the  present  letter  we  have
investigated  DCC  formation in 3+1 dimension. Two scenarios were
considered. In scenario I, we  use  the  equation  of  motion  of
linear $\sigma$ model with quenched initial condition to simulate
DCC  formation  without  friction  and  noise. In scenario II, we
solve the Langevin equation for linear $\sigma$ model,  with  the
same  quenched  initial  condition  to  simulate DCC formation in
presence of friction and noise. Comparative study  of  these  two
scenarios reveal that certainly the noise and friction affect DCC
formation   probability.  However  contrary  to  one  dimensional
calculations, large DCC formation is seen in scenario  II  rather
than  in  scenario  I.  Indeed,  it  will be shown that large DCC
domains are formed in scenario II only, not in scenario  I.  This
effect is particular to heavy ion collisions as explained below.

The  Langevin  equation for linear $\sigma$ model, at temperature
$T$ can be written as,

\begin{equation}
[\frac{\partial^2}{\partial \tau^2} +(\frac{1}{\tau}+\eta_i)
\frac{\partial}{\partial \tau}
-\frac{\partial^2}{\partial x^2} -\frac{\partial^2}{\partial y^2} -
\frac{1}{\tau^2} \frac{\partial^2}{\partial Y^2}
+\lambda (\Phi^2 - f^2_\pi -T^2/2)] \Phi
 = \zeta_i (\tau ,x,y,Y)\label{1}
\end{equation}

\noindent  where  $\Phi_i=(\sigma,\pi_1,\pi_2,\pi_3)$.  $\tau$ is
the  proper  time  and  Y  is  the  rapidity,   the   appropriate
coordinates to describe heavy ion collisions. $\eta_{\sigma,\pi}$
and  $\zeta_{\sigma,\pi}$  are  the friction coefficients and the
white noise for the $\sigma$ and $\pi$ fields. We  note  that  if
the  friction and the noise terms are dropped from eq.\ref{1} the
resulting equation is for the scenario I, i.e. just  the  equation
of motion of linear $\sigma$ model.

The  noise  source  $\zeta$  and  the  friction  $\eta$  are  not
independent. They are  related  by  the  fluctuations-dissipation
relation.  We  use  white  noise  source  with  zero  average and
correlation  as   demanded   from   the   fluctuation-dissipation
relation,

\begin{mathletters}
\begin{eqnarray}
<\zeta_a(\tau,x,y,Y)> =&&0\\
<\zeta_a(\tau,x,y,Y)\zeta_b(\tau^\prime,x^\prime,y^\prime,Y^\prime)>
=&& 2 T \eta \frac{1}{\tau} \delta(\tau-\tau^\prime) \delta(x-x^\prime)
\delta(y-y^\prime)
\delta(Y-Y^\prime) \delta_{ab}
\end{eqnarray}
\label{2}
\end{mathletters}

\noindent where $a,b$ corresponds to $\pi$ or $\sigma$ fields.
A few words are necessary about the use of temperature in the
noise term.  We are approximating DCC, which is a non-equilibrium
phenomena as a equilibrium one. Such an approximation is valid
when the system is not far from equilibrium. Indeed, fluctuation-dissipation
theorem is valid for such a system only.

Friction coefficients is an important ingredient for the Langevin
equation.  In  our  earlier  study,  we  have  assumed  that both
$\sigma$ and  $\pi$  evolve  under  the  influence  of  a  common
friction  \cite{ch99c,ch00}.  In  the  present work we treat them
separately. $\sigma$ and $\pi$ fields evolve under the  influence
of  friction,  appropriate  for  them.  Friction coefficients for
$\sigma$ and $\pi$ have been calculated by Rischke \cite{ri98},

\begin{mathletters}
\begin{eqnarray}
\eta_\pi =&& (\frac{4 \lambda f_\pi}{N})^2 \frac{m_\sigma^2}{4  \pi
m_\pi^3} \sqrt{1-\frac{4 m_\pi^2}{m_\sigma^2}}
\frac{1-exp(-m_\pi/T)}{1-exp(-m_\sigma^2/2 m_\pi T)}
\frac{1}{exp(m_\sigma^2-2m_\pi^2)/2m_\pi T)-1}\\
\eta_\sigma   =&&  (\frac{4  \lambda  f_\pi}{N})^2  \frac{N-1}{8\pi
m_sigma}
\sqrt{1-\frac{4 m_\pi^2}{m_\sigma^2}} \coth \frac{m_\sigma}{4T}
\end{eqnarray} \label{3}
\end{mathletters}

Solution  of  eq.\ref{1}  require  initial fields configurations.
They were distributed according to a random Gaussian with,

\begin{mathletters}
\begin{eqnarray}
<\sigma>=&&(1-f(r))f_\pi \\
<\pi_i>=&&0 \\
<\sigma^2>-<\sigma>^2 = <\pi_i^2>-<\pi_i>^2=   && f_\pi^2/6 f(r)\\
< \dot{\sigma}>=&& <\dot{\pi_i}>=0\\
<\dot{\sigma}^2>=<\dot{\pi}^2>=&& 4 f_\pi^2/6 f(r)
\end{eqnarray}
\label{4}
\end{mathletters}

The interpolation function

\begin{equation}
f(r)=[1+exp(r-r_0)/\Gamma)]^{-1}
\end{equation}

\noindent  separates  the  central  region  from  the rest of the
system. We have  taken  $r_0$=6.4  fm  and  $\Gamma$=.7  fm.  The
initial field configurations corresponds to quench like condition
\cite{ra93}  and  as  told  in  the  beginning are unlikely to be
obtained in a heavy ion collisions. We still use it as  they  are
the  most  favourable  initial  conditions  to  produce  DCC like
phenomena. In the simulation results presented below, the initial
time and  temperature  were  assumed  to  be  $\tau_i$=1fm/c  and
$T_i$=200  MeV.  Effect  of  expansion  was  included through the
cooling law,

\begin{equation}
T(\tau) = T_i \frac{\tau_i}{\tau}
\end{equation}

\noindent which is rather fast for heavy ion collisions. However,
we choose to use it as DCC formation probability increases if the
system cools rapidly.

With  the initial conditions and the cooling law most appropriate
for DCC formation, we have solved the Langevin equation (scenario
II) on a $32^3$ lattice, with lattice spacing of a=1 fm, using  a
time  step of a/10 fm/c. We also use periodic boundary conditions.
The equation of motion for linear $\sigma$ model (scenario I) was
solved similarly, with  identical  initial  conditions.  We  have
continued the evolution of the fields for about 10 fm/c.

We   define   a   correlation   function   at   rapidity  $Y$  as
\cite{asak}

\begin{equation}
C(r,\tau)  =  \frac{  \sum_{i,j} \pi(i) \dot \pi(j)}{\sum_{i,j}
|\pi(i)| |\pi(j)|}
\end{equation}

\noindent  where  the sum is taken over those grid points i and j
such that the distance between i and j is r. In  fig.1,  we  have
compared the correlation function in scenario I and II. Initially
there  is  no  correlation length beyond the lattice spacing of 1
fm. Correlations starts to develop at later  time.  It  increases
for  about  7fm/c,  then  decreases  again. Interestingly, larger
correlation length is  obtained  in  the  scenario  II,  than  in
scenario  I.  Thus at 7 fm/c, correlation length in scenario I is
only $\sim$ 2 fm, while that in  scenario  II  is  $\sim$  4  fm.
Increased  correlation  in  scenario  II  is  contrary to popular
expectation that friction and noise will reduce correlation.

Enhancement  of  correlation length in scenario II, with friction
and noise is corroborated in the pion field distribution also. In
fig.2,  we  have  shown  the  xy  contour  plot  of  the  $\pi_2$
component, at rapidity Y=0. Field distribution at $\tau_i$=1 fm/c
and  after  7  fm/c of evolution are shown. Initially there is no
correlation. Domain like structure is seen both in scenario I and
II, after 7fm/c evolution . The positive and negative  components
of the $\pi_2$ separates out. Here again, much larger domains are
formed  in  scenario  II than in scenario I. It may be noted that
domain like structure seen in one of the component of $\pi$ field
do not  necessarily  convert  into  physical  domains.  They  are
indication  of  larger  correlation  length only. Physical domain
should contain either charged  or  neutral  pion  only.  Thus  in
physical  domain  neutral  to  total  pion  ratio  should  differ
considerably from the isospin symmetric value of 1/3.

Assuming that the pion density is proportional to the amplitude's
square,  in  fig.3, we have shown the contour plot of the neutral
to total pion ratio, at rapidity $Y=0$.

\begin{equation}
R_3(x,y,Y,\tau)=\frac{\pi^2_3}{\pi^2_1+\pi^2_2+\pi^2_3}
\end{equation}

Very  small  or large value of the ratio over an extended spatial
zone   will  be  definite  indication  of  domain  formation.  As
expected,  initially  there  is  no  domain  like  structure.  In
scenario  I, we donot find any large domain like structure in the
ratio  $R_3$  even  after  7  fm/c  evolution. Thus while $\pi_2$
component of the $\pi$ field shows domain  like  structure  at  7
fm/c,  in terms of physical pions there is no domain structure in
scenario I. Large DCC domains are not formed in  linear  $\sigma$
model, even with quenched initial condition and fast cooling law.
In  scenario II however we can find some extended zone with large
or very small value of  the  ratio  $R_3$.  Thus  physical  pions
evolve  into domain like structure in scenario II, with noise and
friction, rather than in scenario I.

Present  simulation  indicate  that  friction and noise do indeed
enhance DCC domain formation possibility. The simulation  results
are contrary to expectation from one dimensional calculations. As
such  noise and friction are supposed to reduce the amplification
of long wavelength modes. We believe  that  present  results  are
particular to heavy ion collisions where proper time and rapidity
are  the  most appropriate coordinate systems. In this coordinate
system, correlation of the noise  decreases  with  (proper)  time
($1/\tau$  dependence,  see eq.\ref{2}). Physically the source of
noise i.e. the medium surrounding the zone where chiral  symmetry
is  restored,  fly  away  with  time,  reducing  the correlation.
However, the friction  coefficients  $\eta_{\sigma,\pi}$  remains
more  or  less  same in the temperature range considered. Thus at
later time evolution of the fields are determined mainly
by the friction.
We also not  that  $\eta_\sigma$  is  considerably  large
($\sim  3 fm^{-1}$). Then once the trajectory enters into unstable
regime, large friction opposes its tendency to come  out  of  the
instability.  Friction  forces  the  trajectory  to  remain in the
unstable regime for longer duration. Indeed, in one dimension, we
have obtained similar result  \cite{ch99a,ch99b}.  3d  simulation
confirms our results in one dimension.

Though  noise and friction enhances the DCC formation probability,
it may not be easy to detect it. In  fig.4,  we  have  shown  the
rapidity distribution of the neutral to pion ratio,

\begin{equation}
\frac{N_{\pi_0}}{N_{\pi_{total}}}=\frac{\int  \pi^2_3 dx dy}{\int
(\pi^2_1+\pi^2_2+\pi^2_3) dx dy}
\end{equation}

\noindent  at  different time intervals. Through out the range of
rapidity, the ratio fluctuate about its isospin  symmetric  value
of  1/3.  As expected fluctuations are larger in scenario II than
in scenario I. However, if we remember that the pions  that  will
be detected are integrated over time, then it is obvious that the
fluctuations   will   be   considerably   less.   Thus   even  in
event-by-event analysis, it will be difficult to tell  about  DCC
formation  from  rapidity  distribution  of pions only. Much more
study is needed to resolve the issue.

In  summary,  we  have  considered  disoriented chiral condensate
domain formation in two scenarios, one without noise and friction
(scenario I) and the other with noise and friction (scenario II).
In scenario I, equation of motion for the linear  $\sigma$  model
fields  and  in  scenario  II,  the Langevin equations for linear
$\sigma$ model were solved. Using the most ideal  conditions  for
DCC  formation, i.e. quenched initial condition and fast cooling,
we have evolved the fields for 10 fm/c. It was seen that in  both
the  scenario,  $\pi  \pi$ correlation increases with time till 7
fm/c, it  then  decreases.  Larger  correlation  is  obtained  in
scenario II than in scenario I. Evidence of increased correlation
is  also  obtained  from  the contour plot of pion field. $\pi_2$
component of the pion field shows domain like structure. Positive
and negative component of fields separates out at late time. Here
again, domain like structure is more  prominent  in  scenario  II
than  in  scenario  I.  Contour plot of the neutral to total pion
ratio suggest that though domain like structure is  seen  in  one
component of the $\pi$ field, there is no (large) physical domain
formation  in scenario I. On the contrary, large physical domains
are seen to be formed in scenario II after 7 fm/c  of  evolution.
We  have also studied the rapidity distribution of the neutral to
pion ratio in both the scenarios. Throughout the rapidity  range,
the  ratio  fluctuate  about  the isospin symmetric value of 1/3,
fluctuations being more  in  scenario  II  than  in  scenario  I.
However, it was also noted that fluctuations will be considerably
less  when integrated over time. Rapidity distribution of neutral
to total pion ratio may not be able to single out the DCC events.

\eject
\begin{figure}
\centerline{\psfig{figure=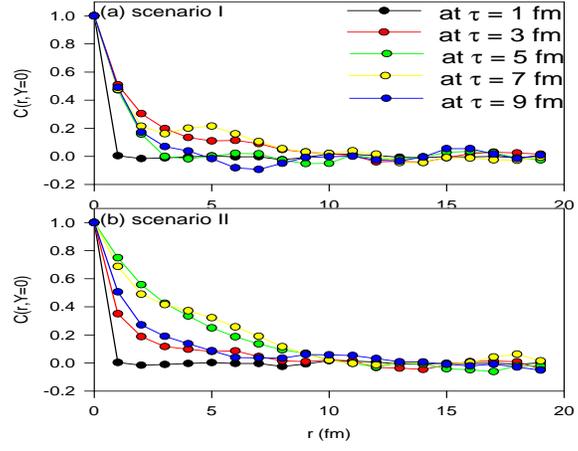,height=10cm,width=10cm}}
\caption{Evolution  of the pion correlation fucntion  with  time,
in scenario I and II.}
\end{figure}

\begin{figure}
\centerline{\psfig{figure=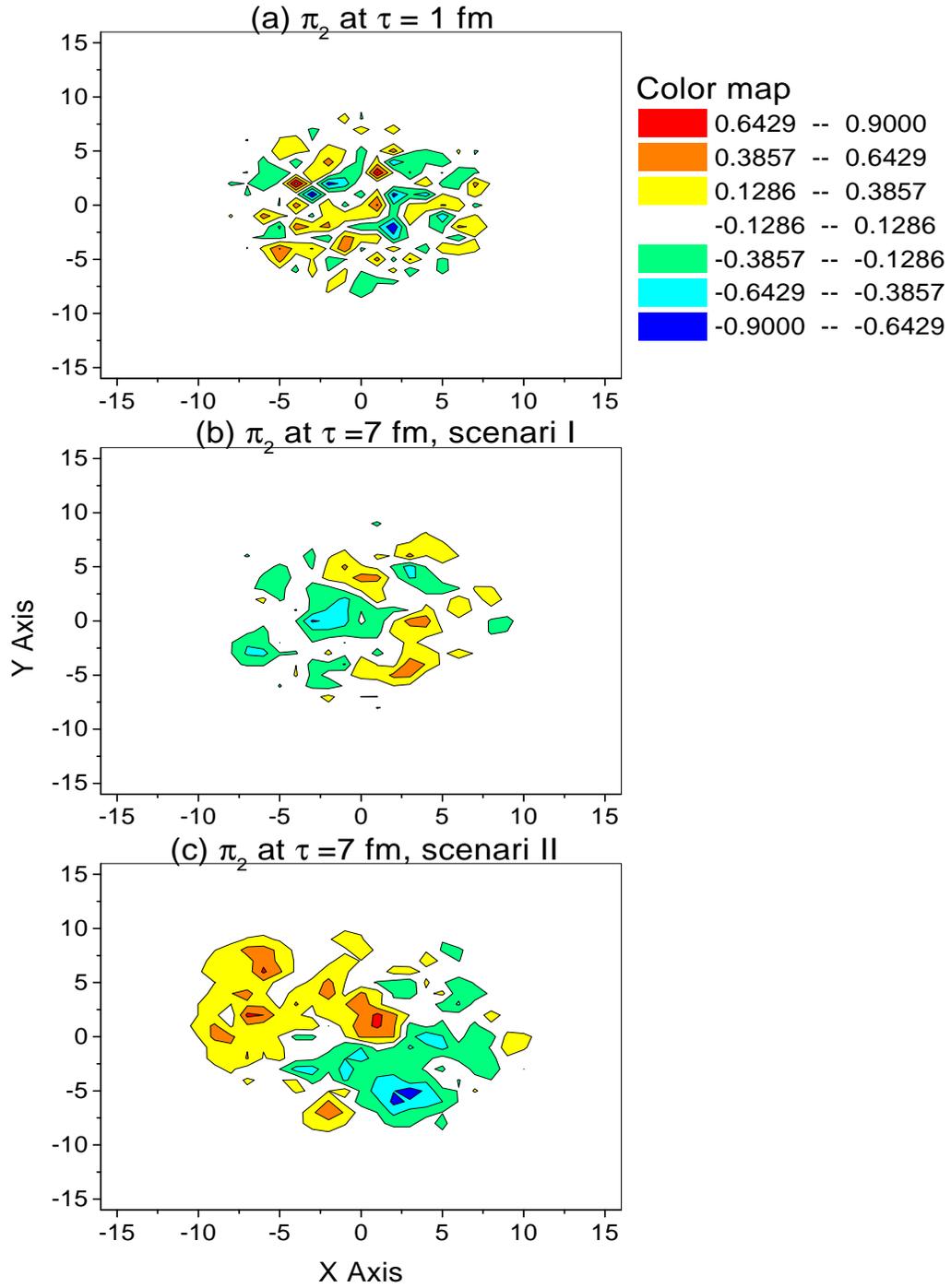,height=20cm,width=15cm}}
\caption{Contour  plot of the $\pi_2$ component of the pion field
at rapidity Y=0,  at  initial  time  $\tau_i$=1  fm/c  and  after
evolution for 7 fm/c.}
\end{figure}

\begin{figure}
\centerline{\psfig{figure=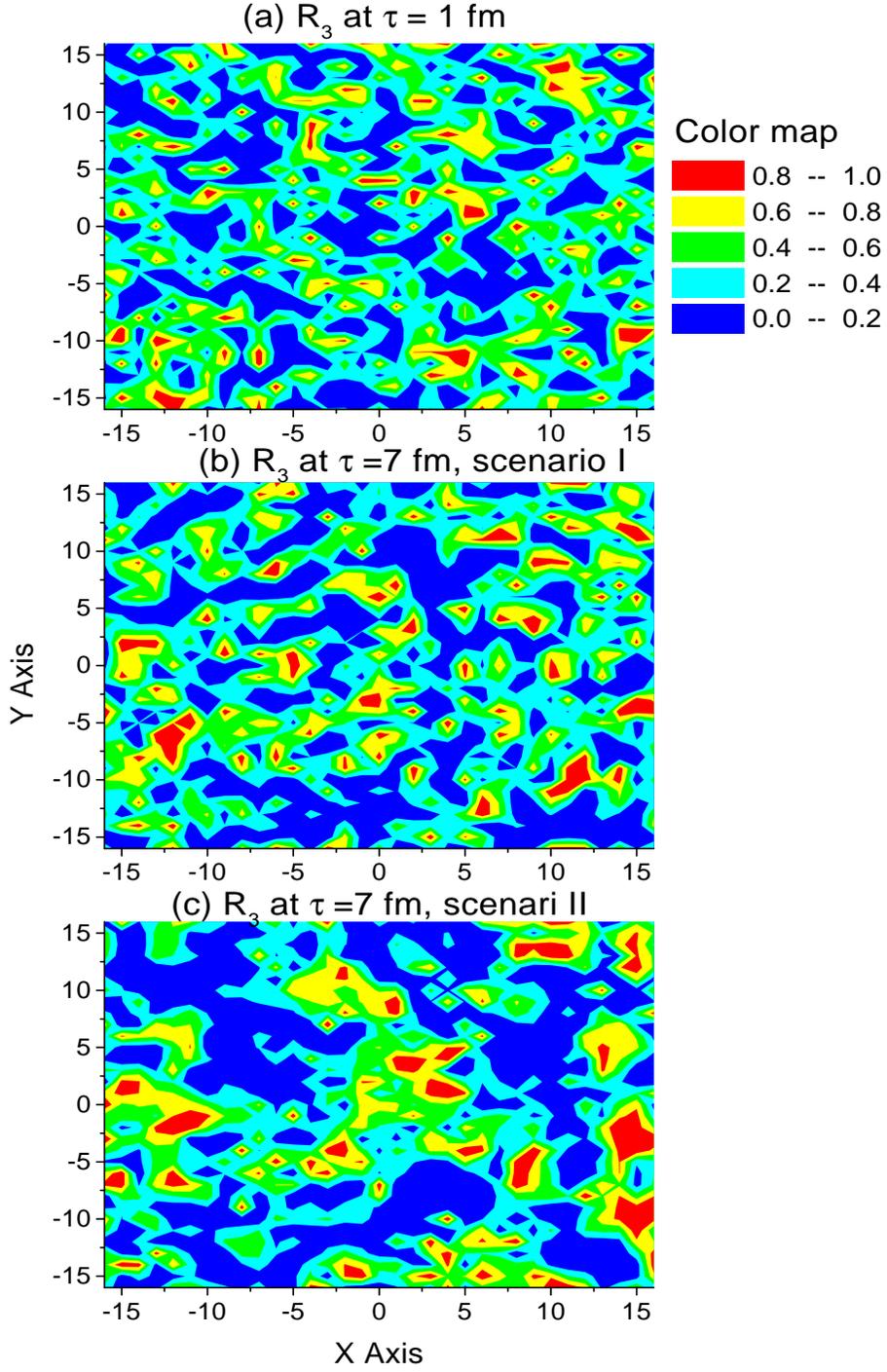,height=20cm,width=15cm}}
\caption{Contour  plot  of  the  neutral  to total pion ratio, at
rapidity Y=0, at initial time $\tau_i$=1 fm/c and after evolution
for 7 fm/c.}
\end{figure}

\begin{figure}
\centerline{\psfig{figure=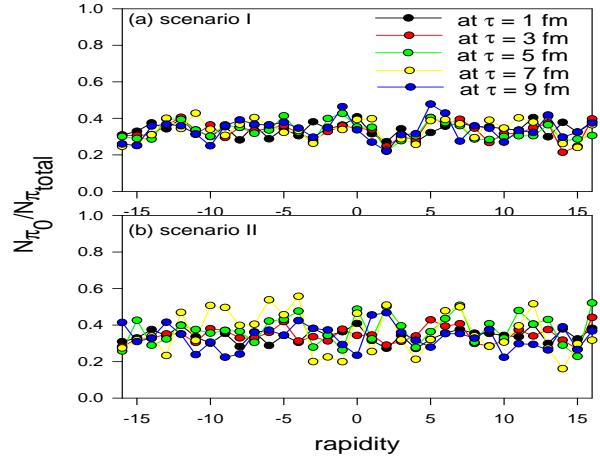,height=10cm,width=10cm}}
\caption{Rapidity distribution of neutral to total pion ratio, in
scenario  I  and  II.  The  ratio at different time intervals are
shown.}
\end{figure}
\end{document}